\documentclass[iop]{emulateapj}

\shorttitle{ASTEROIDAL ACCRETION AT WHITE DWARFS}
\shortauthors{Farihi et al.}

\begin{document}

\title{STRENGTHENING THE CASE FOR ASTEROIDAL ACCRETION:\\
EVIDENCE FOR SUBTLE AND DIVERSE DISKS AT WHITE DWARFS}

\author{J. Farihi\altaffilmark{1},
			M. Jura\altaffilmark{2},
			J.-E. Lee\altaffilmark{3,4},
	 		B. Zuckerman\altaffilmark{2}}

\altaffiltext{1}{Department of Physics \& Astronomy,
			University of Leicester,
			Leicester LE1 7RH, UK; 
			jf123@star.le.ac.uk}

\altaffiltext{2}{Department of Physics \& Astronomy,
			University of California,
			Los Angeles, CA 90095; 
			jura,ben@astro.ucla.edu}
			
\altaffiltext{3}{Department of Astronomy \& Space Science,
			Astrophysical Research Center for the Structure and Evolution of the Cosmos,
			Sejong University, 
			Seoul 143-747, Korea; 
			jelee@sejong.ac.kr}
			
\altaffiltext{4}{English translation of \includegraphics[width=1.0cm]{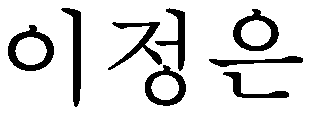}}

\begin{abstract}

{\em Spitzer Space Telescope} IRAC $3-8\,\mu$m and {\em AKARI} IRC $2-4\,\mu$m photometry 
are reported for ten white dwarfs with photospheric heavy elements; nine relatively cool stars with 
photospheric calcium, and one hotter star with a peculiar high carbon abundance.  A substantial 
infrared excess is detected at HE\,2221$-$1630, while modest excess emissions are identified at 
HE\,0106$-$3253 and HE\,0307$+$0746, implying these latter two stars have relatively narrow 
($\Delta r < 0.1\,R_{\odot}$) rings of circumstellar dust.  A likely 7.9\,$\mu$m excess is found at 
PG\,1225$-$079 and may represent, together with G166-58, a sub-class of dust ring with a large 
inner hole.  The existence of attenuated disks at white dwarfs substantiates the connection between 
their photospheric heavy elements and the accretion of disrupted minor planets, indicating many 
polluted white dwarfs may harbor orbiting dust, even those lacking an obvious infrared excess.
\end{abstract}


\section{INTRODUCTION}

The large-scale composition of extrasolar minor planets can be indirectly measured 
using white dwarfs as astrophysical detectors \citep{jur09b,zuc07}.  Rings of warm dust 
(and in some cases gaseous debris) revealed at over one dozen white dwarfs \citep{far09a} 
are situated within the Roche limits \citep{gan06} of their respective stellar hosts, and likely 
originated via the tidal disruption of minor (or possibly major) planets \citep{jur03}.  These 
closely orbiting disks rain metals onto the stellar photosphere, and contaminate an otherwise
pristine hydrogen or helium atmospheric composition.  For any particular cool white dwarf, the
timescales for individual heavy elements to sink below the high gravity photosphere differ by 
less than a factor of three, yet are always orders of magnitude shorter than the cooling age 
\citep{koe09a}.  With an appropriate treatment of the accretion history, photospheric heavy 
element abundances can be used as a measure of the composition of the tidally destroyed, 
polluting parent body or bodies \citep{kle10,jur08}.

The spectacularly debris-polluted white dwarf GD\,362 is an excellent example of this 
potential.  Its circumstellar disk re-emits at least 3\% of the incident stellar radiation; one-third 
of this is carried by a strong $10\,\mu$m silicate emission feature \citep{jur07b}.  The star itself 
is polluted by at least 15 elements heavier than hydrogen and helium, in an arrangement that 
is rich in refractory elements and deficient in volatiles; a pattern that broadly mimics the inner 
Solar system, and comparable with the outer composition of the Earth and Moon \citep{zuc07}.  
A recent analysis of the unusual atmospheric mix of hydrogen and helium, together with X-ray 
constraints on the current hydrogen accretion rate at GD\,362, suggests that if a single parent 
body gave rise to both the currently observed disk and the panoply of heavy elements in its 
atmosphere, then its mass would be larger than Callisto, and possibly larger than Mars
\citep{jur09b}.

Observations of a large number and variety of metal-polluted white dwarfs are necessary to 
better understand and constrain the connection between rocky extrasolar parent bodies, the 
circumstellar dust and gas, and the subsequent photospheric pollutions they create in white 
dwarfs \citep{zuc03}.  The lifetimes of the disks at white dwarfs are poorly constrained at 
present, though there is some indication their lifetimes do not significantly exceed $10^5$\,yr 
\citep{far09a,kil08}.  Another outstanding issue is the nature of the large metal abundances 
in stars lacking excess infrared emission; tenuous, gaseous, or previously-accreted disks 
are strong possibilities.  While the evolution of optically thick dust at white dwarfs should be 
dominated by viscous dissipation-spreading, to date only G166-58 stands out as a possible 
indicator of how competing mechanisms, such as additional impacts or gas drag at the inner 
disk edge, may play a role \citep{jur08,far08}.

{\em Spitzer} IRAC and {\em AKARI} IRC photometric imaging observations are presented for 
nine metal-polluted white dwarfs chosen for their high calcium abundances and relatively short 
cooling ages.  The observations are presented in \S2, the results in \S3, and the implications for 
the minor planet accretion hypothesis and future dust searches are discussed in \S4.

\begin{deluxetable*}{ccccccc}
\tabletypesize{\footnotesize}
\tablecaption{{\em Spitzer} and {\em AKARI} \ White Dwarf Targets\label{tbl1}}
\tablewidth{0pt}
\tablehead{
\colhead{WD}					&
\colhead{Name}				&
\colhead{SpT}					&
\colhead{$V$}					&
\colhead{[Ca/H(e)]}				&	
\colhead{Telescope}				&
\colhead{Refs}					\\
&
&
&(mag)
&
&	
&}

\startdata

0047$+$190\tablenotemark{a}	&HS\,0047$+$1903	&DAZ	&16.1	&$-6.1$		&{\em Spitzer}			&1\\
0106$-$328\tablenotemark{a}	&HE\,0106$-$3253	&DAZ	&15.5	&$-5.8$		&{\em Spitzer}			&1\\
0307$+$077\tablenotemark{a}	&HS\,0307$+$0746	&DAZ	&16.4	&$-7.1$		&{\em Spitzer}			&1\\
0842$+$231\tablenotemark{a}	&Ton\,345 		&DBZ	&15.9	&$-6.9$		&{\em AKARI}			&2\\
1011$+$570				&GD\,303			&DBZ	&14.6	&$-7.8$		&{\em Spitzer / AKARI}	&3\\
1225$-$079				&PG				&DZAB	&14.8	&$-7.9$		&{\em Spitzer}			&3,4\\
1542$+$182				&GD\,190			&DBQ	&14.7	&\nodata		&{\em Spitzer}			&5\\
1709$+$230				&GD\,205			&DBAZ	&14.9	&$-8.0$		&{\em Spitzer / AKARI}	&4,6\\
2221$-$165\tablenotemark{a}	&HE	2221$-$1630	&DAZ	&16.1	&$-7.2$		&{\em Spitzer}			&1\\
2229$+$235\tablenotemark{a}	&HS	2229$+$2335	&DAZ	&15.9	&$-5.9$		&{\em Spitzer}			&1

\enddata

\tablerefs{
1) \citealt{koe06}
2) \citealt{gan08}
3) \citealt{wol02}
4) \citealt{koe05}
5) \citealt{pet05}
6) \citealt{vos07}}

\tablenotetext{a}{The WD numbers for these stars are unofficial designations, but correctly reflect 
the conventional use of epoch B1950 coordinates.  Abundances are expressed as  [X/Y] = $\log\,[
n({\rm X})/n({\rm Y})]$.\\}

\end{deluxetable*}

\section{OBSERVATIONS AND DATA}

The target sample of stars was chosen based on the prior result that nine of 14 white dwarfs 
with inferred metal accretion rates above $3\times10^8$\,g\,s$^{-1}$ and cooling ages less 
than 1.0\,Gyr have infrared excess \citep{far09a}.  All nine metal-contaminated white dwarfs 
in Table \ref{tbl1} satisfy this criteria; five DAZ stars with high rates of ongoing metal accretion, 
plus three DBZ stars with high calcium abundances.  Table \ref{tbl2} lists the relevant parameters 
for the DAZ and DBZ stars.  Also included as a target is the 22\,000\,K DBQ white dwarf GD\,190, 
with an anomalous atmospheric carbon abundance \citep{pet05}.  Ton\,345 was included based 
on the presence of circumstellar, gaseous metal emission \citep{gan08}.

\begin{figure*}
\epsscale{1.1}
\plotone{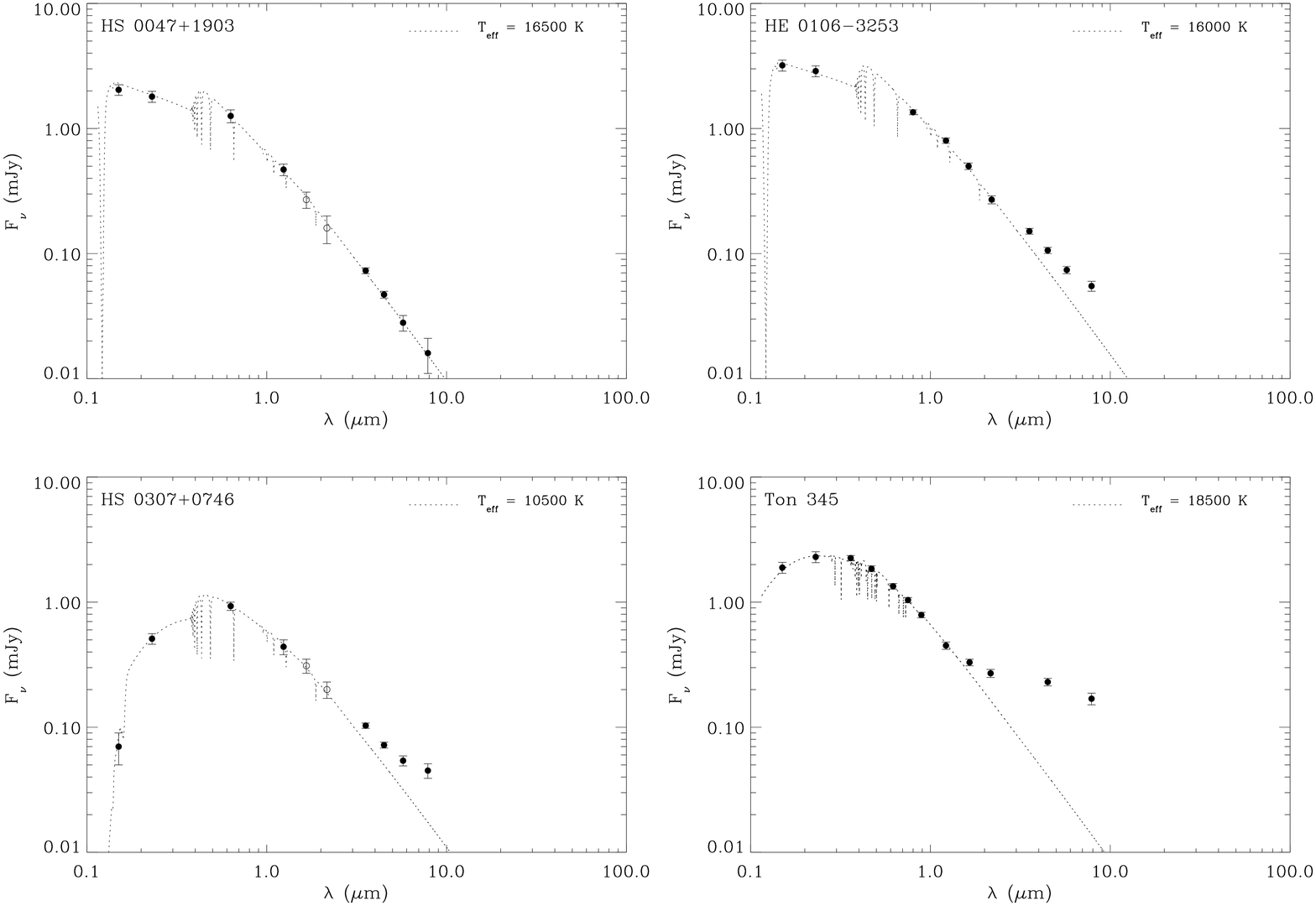}
\caption{SEDs of HS\,0047$+$1903, HE\,0106$-$3253,HS\,0307$+$0746, and 
Ton\,345 fitted with DA and DB white dwarf models \citep{koe09b}.  Filled circles represent 
the photometry \citep{mel10,bri10,far09b,aba09,skr06,cmc06,den05,mar05} while open circles 
are fluxes based on spectroscopically derived colors \citep{kil06}, and hence not independent 
measurements.\\
\label{fig1}}
\epsscale{1.0}
\end{figure*}

\subsection{{\em Spitzer} IRAC Observations}

Mid-infrared imaging observations of the white dwarf targets were obtained with the 
{\em Spitzer Space Telescope} \citep{wer04} during Cycle 5 using the Infrared Array Camera 
(IRAC; \citealt{faz04}) at all four wavelengths, namely 3.6, 4.5, 5.7, and 7.9\,$\mu$m.  The total 
integration time in each channel was 600\,s, where the observations consisted of 20 frames 
taken in the cycling (medium) dither pattern with 30\,s individual exposures.  All images were 
analyzed, including photometry and upper limits, as in \citet{far09a} using both $1\farcs2$ 
and $0\farcs6$\,pixel$^{-1}$ mosaics created using MOPEX.

In cases where the flux of a neighboring source was a potential contaminant of the white dwarf 
photometry, steps were taken to minimize or remove any such external contributions, including 
small aperture radii and point-spread function (PSF) fitting with {\sf daophot}.  To this end, the 
newly available (beginning with pipeline version S18.5) $0\farcs6$\,pixel$^{-1}$ mosaics were 
useful to better constrain and eliminate sources of photometric confusion.  While the IRAC Data 
Handbook \citep{ssc06} provides information necessary to perform photometry on point sources 
in the native $1\farcs2$\,pixel$^{-1}$ scale images, the provided aperture corrections do not 
take advantage of the available smaller pixel scale.  A comparison of identical observation sets, 
processed into both $1\farcs2$ and $0\farcs6$\,pixel$^{-1}$ mosaics, reveals a modest but clear 
increase (around 20\% by full width at half maximum) in spatial resolution.  This information gain 
at the smaller pixel scale is most evident at the two shortest wavelengths channels, where the 
diffraction limited PSF of the telescope is significantly under-sampled \citep{ssc06}.

\begin{deluxetable}{ccccc}
\tabletypesize{\footnotesize}
\tablecaption{{\em Spitzer} IRAC Fluxes and Upper Limits\label{tbl2}}
\tablewidth{0pt}
\tablehead{
\colhead{WD}					&
\colhead{$F_{3.6\mu{\rm m}}$}		&
\colhead{$F_{4.5\mu{\rm m}}$}		&
\colhead{$F_{5.7\mu{\rm m}}$}		&
\colhead{$F_{7.9\mu{\rm m}}$}		\\	
&($\mu$Jy)	
&($\mu$Jy)		
&($\mu$Jy)	
&($\mu$Jy)}

\startdata

0047$+$190			&$73\pm4$		&$47\pm3$		&$28\pm4$		&$16\pm5$\\
0106$-$328			&$151\pm8$		&$106\pm6$		&$74\pm5$		&$55\pm5$\\
0307$+$077			&$103\pm5$		&$72\pm4$		&$54\pm5$		&$45\pm6$\\
1011$+$570			&$220\pm11$		&$137\pm7$		&$86\pm6$		&$55\pm5$\\
1225$-$079			&$328\pm16$		&$224\pm11$		&$139\pm8$		&$106\pm8$\\
1542$+$182			&$194\pm10$		&$121\pm6$		&$75\pm6$		&$38\pm5$\\
1709$+$230			&$180\pm9$		&$107\pm6$		&$63\pm5$		&$32\pm5$\\
2221$-$165			&$204\pm10$		&$172\pm9$		&$157\pm9$		&$119\pm8$\\
2229$+$235			&$61\pm6$		&$37\pm4$		&$20\pm5$		&18\tablenotemark{a}

\enddata

\tablenotetext{a}{$3\sigma$ upper limit.}

\end{deluxetable}

\subsection{{\em AKARI} IRC Observations}

Observations with {\em AKARI} \citep{mur07} were executed using the Infrared Camera 
(IRC; \citealt{ona07}) in the N2, N3 and N4 filters (2.34, 3.19, and 4.33\,$\mu$m) for three 
white dwarfs listed in Table \ref{tbl1}.  These stars were imaged in 2008 November and 2009 
March-April using the Astronomical Observational Template IRCZ3, which was designed for 
imaging with the N2, N3, and N4 filters in a pointed observation.  The basic data reduction, 
including dark subtraction, linearity correction, distortion correction, flat fielding, image stacking, 
and absolute position determination were performed by the standard IRC Imaging Data 
Reduction Pipeline for Phase 3 data (version 080924P3).  The IRAF package {\sf daophot} 
was used for photometry, with an aperture radius of 5 pixels ($7\farcs3$) and corrected to 
the standard 10 pixel radius using aperture corrections derived from bright sources within 
individual images.  Flux calibration was achieved using conversion factors provided by the 
IRC team\footnote{http://www.ir.isas.jaxa.jp/ASTRO-F/Observation/DataReduction/IRC/ConversionFactor\_\_090824.html}.

\begin{deluxetable}{cccc}
\tabletypesize{\footnotesize}
\tablecaption{{\em AKARI} IRC Fluxes\label{tbl3}}
\tablewidth{0pt}
\tablehead{
\colhead{WD}					&
\colhead{$F_{2.3\mu{\rm m}}$}		&
\colhead{$F_{3.2\mu{\rm m}}$}		&
\colhead{$F_{4.3\mu{\rm m}}$}		\\
&($\mu$Jy)	
&($\mu$Jy)		
&($\mu$Jy)}

\startdata

0842$+$231			&$293\pm22$		&$271\pm23$		&$218\pm18$\\
1011$+$570			&$586\pm46$		&$289\pm22$		&$146\pm12$\\
1709$+$230			&$418\pm32$		&$237\pm18$		&$111\pm11$

\enddata

\end{deluxetable}

\section{ANALYSIS AND RESULTS}

\begin{figure*}
\epsscale{1.1}
\plotone{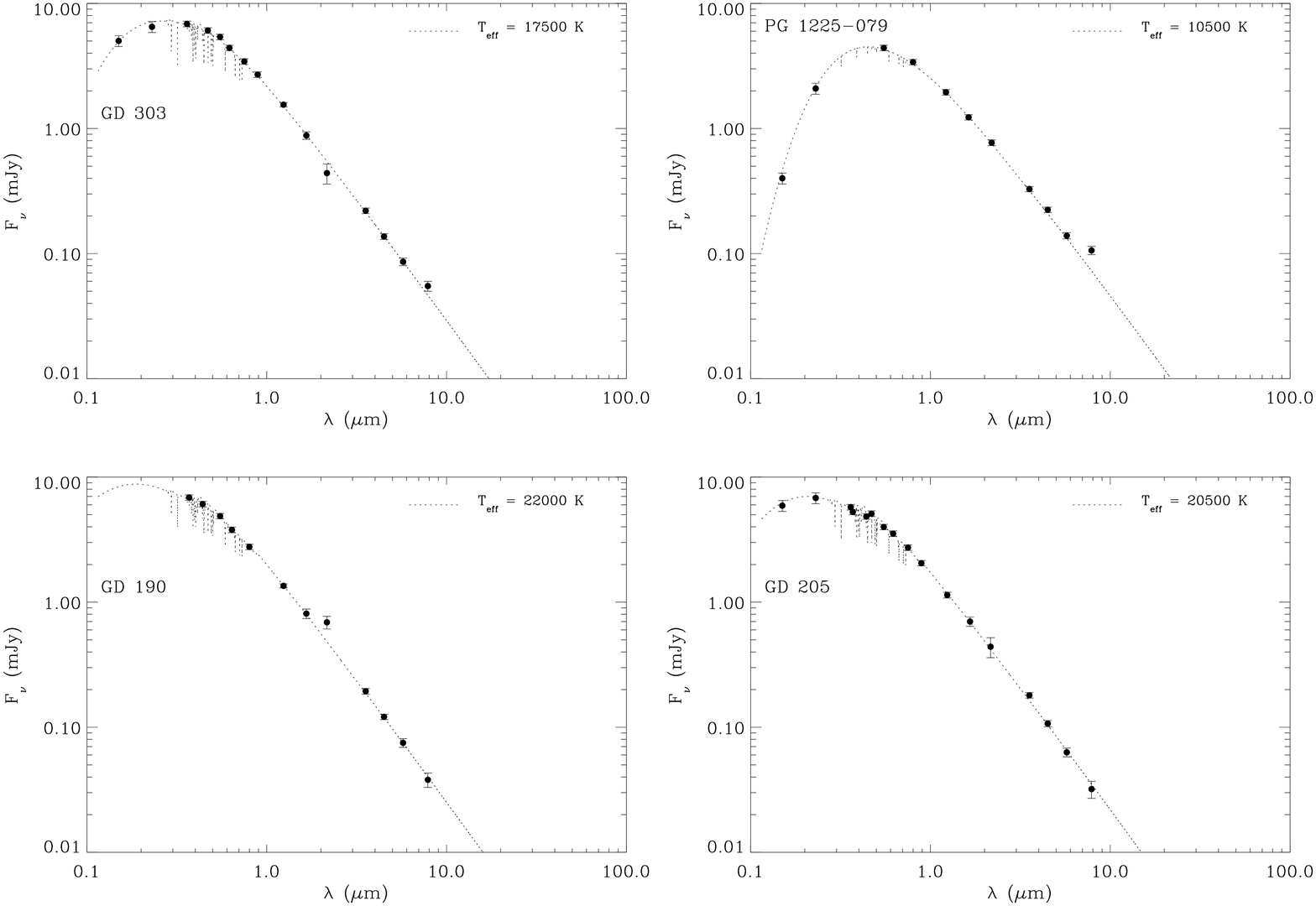}
\caption{SEDs of GD\,303, PG\,1225$-$079, GD\,190, and GD\,205 fitted DB white dwarf 
models \citep{koe09b}.  Filled circles represent the photometry \citep{far09b,aba09,lan07,
skr06,den05,mar05,mcc99}.  The apparent 2MASS $K_s$-band excess at GD\,190 is spurious.
\label{fig2}}
\epsscale{1.0}
\end{figure*}

Tables \ref{tbl2}--\ref{tbl3} list the flux determinations and established upper limits for the 
science targets.  Figures \ref{fig1}$-$\ref{fig3} plot the measured mid-infrared fluxes together 
with available near-infrared and optical data \citep{aba09,far09b,skr06,mar05, mcc08,lan07,
kil06,cmc06,sts06,zac05,den05,mon03}.  For a few of the metal-rich target stars, \citet{kil06} has 
published spectroscopically derived $J-H$ and $H-K$ colors, such that the $H$- and $K$-band 
fluxes of these stars are anchored to the 2MASS $J$-band flux, and hinge upon its accuracy.
IRTF $JHK$ photometry of HE\,0106$-$3252 and PG\,1225$-$079 are taken from \citet{far09b}.  
Where available, {\em Galaxy Evolution Explorer} ({\em GALEX}; \citealt{mar05}) far- and 
near-ultraviolet photometric fluxes are plotted to assist in establishing the appropriate brightness 
level of the white dwarf.  However, because the interstellar extinction to these stars is not known, 
the {\em GALEX} fluxes were used with some caution.  

Five of the target white dwarfs come from the Hamburg Schmidt (HS; \citealt{hag95}) and 
the Hamburg ESO Schmidt (HE; \citealt{wis96}) quasar surveys.  The coordinates in SIMBAD 
for these five white dwarfs have poor precision, and are often missing from the white dwarf 
catalog \citet{mcc99,mcc08}.  The correct positions for these white dwarfs were provided by 
the SPY team (R. Napiwotzki 2007, private communication), and their coordinates in the 
IRAC images are given in Table \ref{tbl4}.  The HE and HS white dwarfs tend to suffer from 
a lack of accurate ground-based photometry in the literature.  A typical star has entries in a 
few photographic catalogs (30\% errors), variable quality $r'$-band CCD photometry from 
the Carlsberg Meridian Catalog (CMC; \citealt{cmc06}), and a DENIS $I$-band measurement
if in the southern hemisphere.  For HS\,0047$+$0093, HS\,0307$+$0746, HE\,2221$-$1630, 
and HS\,2229$+$2335, the only near-infrared photometry is the 2MASS $J$-band value, as 
the $H$- and $K_s$-band data are substantially less reliable owing to the faintness of these 
stars \citep{skr06}.

\begin{deluxetable}{ccc}
\tabletypesize{\footnotesize}
\tablecaption{IRAC Coordinates for HE\,and HS\,White Dwarfs \label{tbl4}}
\tablewidth{0pt}
\tablehead{
\colhead{Star}				&
\colhead{$\alpha$}			&
\colhead{$\delta$}}

\startdata

HS\,0047$+$1903		&$00^{\rm h} 50^{\rm m} 12.47^{\rm s}$	&$+19\arcdeg 19' 48\farcs2$\\
HE\,0106$-$3253		&$01^{\rm h} 08^{\rm m} 36.04^{\rm s}$	&$-32\arcdeg 37' 43\farcs5$\\		
HS\,0307$+$0746		&$03^{\rm h} 10^{\rm m} 09.12^{\rm s}$	&$+07\arcdeg 57' 31\farcs5$\\
HE\,2221$-$1630		&$22^{\rm h} 24^{\rm m} 17.44^{\rm s}$	&$-16\arcdeg 15' 47\farcs7$\\
HS\,2229$+$2335		&$22^{\rm h} 31^{\rm m} 45.49^{\rm s}$	&$+23\arcdeg 51' 24\farcs1$

\enddata

\tablecomments{As measured on the IRAC array based on image header astrometry; the 
epochs all correspond to Cycle 5 dates between 2008.9 and 2009.3 (equinox 2000).\\}

\end{deluxetable}

All available photometric data for each white dwarf were fitted by a pure hydrogen or pure 
helium white dwarf model with $\log\,[g\,({\rm cm\,s}^{-2})]=8.0$ and effective temperature 
to the nearest 500\,K \citep{koe09b}.  Because the surface gravity and the atmospheric 
composition are fixed, and because the effective temperature resolution is limited, the 
models shown in the figures do not necessarily describe accurately or strongly constrain 
the effective temperature of the stars.  Rather, the atmospheric model fits predict the expected 
level of mid-infrared photospheric emission, and thereby constrain the presence or absence 
of infrared excess \citep{far09b}.  Relative to the {\em Spitzer} data, the {\em AKARI} images 
were found to be shallow, with patchy sky background, and highly asymmetric stellar images.  
For this reason, it is likely the errors used here are underestimated, and these data were given 
little weight in the final analysis.

\subsection{Stars With Infrared Excess}

\begin{deluxetable}{cccccc}
\tabletypesize{\footnotesize}
\tablecaption{Flux Determinations\tablenotemark{1} for HE\,0106$-$3253\label{tbl5}}
\tablewidth{0pt}
\tablehead{
\colhead{Method\tablenotemark{2}}	&
\colhead{$F_{3.6\mu{\rm m}}$}		&
\colhead{$F_{4.5\mu{\rm m}}$}		&
\colhead{$F_{5.7\mu{\rm m}}$}		&
\colhead{$F_{7.9\mu{\rm m}}$}		\\	
&($\mu$Jy)	
&($\mu$Jy)		
&($\mu$Jy)	
&($\mu$Jy)}

\startdata

a		&$154.4\pm10.9$		&$106.6\pm8.3$		&$75.3\pm7.6$		&$56.1\pm7.7$\\
b		&$158.0\pm11.2$		&$108.0\pm8.4$		&$77.4\pm7.8$		&$56.1\pm7.7$\\
c		&$147.1\pm10.4$		&$106.4\pm7.6$		&$72.7\pm5.9$		&$54.0\pm5.5$
						
\enddata

\tablenotetext{1}{Adopted fluxes are the weighted average of methods a and c.}

\tablenotetext{2}{Methods: a) PSF fitting of both white dwarf and neighboring source; b) PSF 
fitting of white dwarf only; c) Relative aperture photometry at $r=1\farcs2$ using nearby star 
(\S 3.1).}

\end{deluxetable}

\begin{figure}
\epsscale{1.2}
\plotone{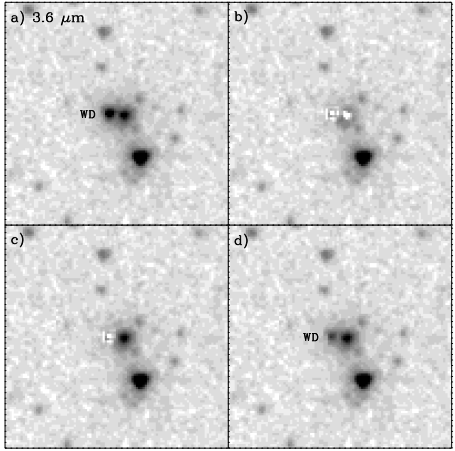}
\caption{IRAC 3.6\,$\mu$m mosaics of HE\,0106$-$3253: a) white dwarf and nearby source; 
b) both sources fitted and subtracted with {\sf daophot}; c) white dwarf removed by {\sf daophot}; 
d) expected white dwarf photospheric flux removed by {\sf daophot}, revealing a point-like 
excess.  Each panel is $1'$ on a side where upwards corresponds to position angle 
$36.0\arcdeg$.
\label{fig4}}
\epsscale{1.0}
\end{figure}

{\em HE\,0106$-$3253}.  The IRAC images of this star reveal another source separated by
$4\farcs4$ at position angle $299\arcdeg$, as measured by {\sf daophot} using the $0\farcs6$ 
pixel$^{-1}$ images (Figure \ref{fig4}).  The neighboring source appears to be a background 
galaxy based on its half maximum intensity at 3.6 and 4.5\,$\mu$m, measured in a direction 
orthogonal to its position angle relative to the white dwarf.  Two independent methods were 
used to measure the flux of the white dwarf without contamination from the nearby galaxy.  
First, {\sf daophot} was employed to fit both the white dwarf and the nearby galaxy (thus 
treating it as a point source) simultaneously.  This was done with two independent sets of
PSF stars, one derived from archival IRAC images taken for the Galactic component of the 
First Look Survey, and the other created using a few to several of bright point sources within 
the science target images.  Additionally, {\sf daophot} was run to fit the white dwarf only, with 
no information provided on the neighboring source.  It is noteworthy that the task returned 
similar results for these two approaches, within 3\% at all four IRAC wavelengths.

\begin{figure}
\epsscale{1.2}
\plotone{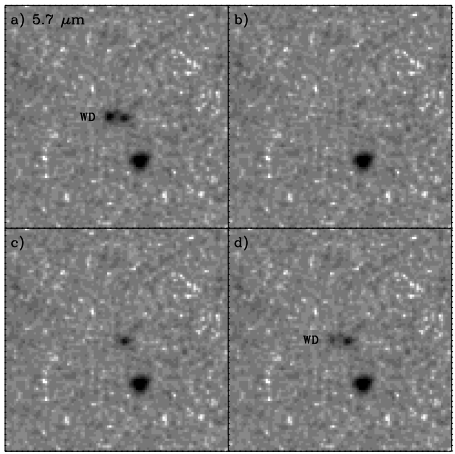}
\caption{Same as Figure 5, but for 5.7\,$\mu$m.  Again, excess flux is revealed after removal of 
the expected white dwarf photosphere.\\
\label{fig5}}
\epsscale{1.0}
\end{figure}

Second, relative aperture photometry was performed on HE\,0106$-$3253 and a moderately
bright, nearby star 2MASS\,J01083497$-$3237486; this star can be seen in Figures \ref{fig4} 
and \ref{fig5}, $33\farcs6$ distant from the white dwarf.  Robust centroids were determined for 
each star using both gaussian profile analysis and {\sf daophot} PSF fitting; the centroids in all 
four channels were averaged for each star and found to agree to within 1 standard deviation
of 0.10\,pixels.  The centroid for the comparison star was taken to be fixed while photometry for
the white dwarf was performed at a total of nine positions; at the nominal centroid plus at eight 
positions along a circle of radius 0.10\,pixels about this center, spaced evenly at angles 45 to 
$360\arcdeg$ in the $xy$ image plane.  Photometry was executed in half-pixel steps from 1 to 
4\,pixels.  The same relative photometric analysis was also performed for two randomly selected 
point sources (called 'star1' and 'star2') within the HE\,0106$-$3253 IRAC field of view.  Lastly, 
the 2MASS comparison star was input to {\sf daophot} as a reference PSF, and the white dwarf 
flux was re-measured as above.

\begin{figure*}
\epsscale{1.1}
\plotone{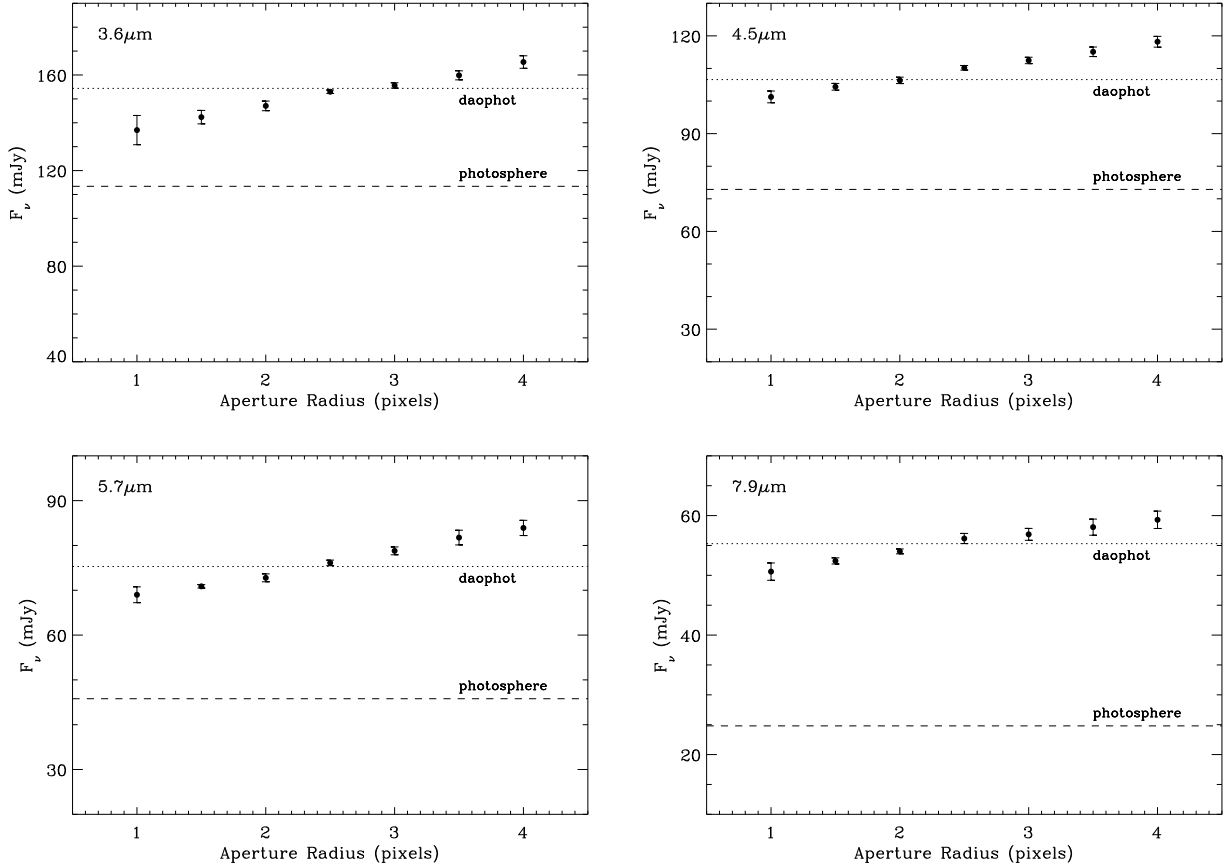}
\caption{IRAC fluxes for HE\,0106$-$3253 derived by photometry relative to the nearby 
comparison star 2MASS\,J01083497$-$3237486 as a function of aperture radius.  The data 
points and error bars represent the average and $3\sigma$ deviations for an ensemble of nine 
white dwarf centroids.  The dotted line marks the white dwarf flux as determined by {\sf daophot}, 
while the dashed line shows the expected photospheric flux.\\
\label{fig6}}
\epsscale{1.0}
\end{figure*}

Figure \ref{fig6} plots the results of the relative flux measurements between the white 
dwarf and 2MASS comparison star.  Each datum is the average and $3\sigma$ error among 
the nine positions where the white dwarf flux was measured.  Also shown in the plot are the 
fluxes as determined by {\sf daophot}, averaged over the 3 sets of PSF stars, and the expected 
white dwarf photospheric fluxes as determined by the model shown in Figure \ref{fig1}.  Two 
things are apparent from the figure and analysis:  1) HE\,0106$-$3253 is photometrically 
extended beyond $r=2$\,pixels, and possibly beyond 1.5\,pixels; 2) the {\sf daophot} flux 
measurements are in good agreement with the fluxes determined by relative photometry 
with the 2MASS comparison star at $r=2$\,pixels.

The fluxes determined by these methods are listed in Table \ref{tbl5}, and a series of 
representative images are shown in Figures \ref{fig4} and \ref{fig5}.  Both methods yield 
IRAC fluxes in excess of the predicted photospheric emission of the white dwarf based on the 
model plotted in Figure \ref{fig1}; the weighted average of these is plotted in the Figure and 
listed in Table \ref{tbl2}.  The excess at each IRAC wavelength lies in the range $3-5\sigma$ 
using the errors derived via the individual measurements, or within $5-6\sigma$ using the 
weighted average errors.  The nearby galaxy cannot account for the results of the PSF fitting 
photometry where extracted sources must conform to a relatively static, model point source.  
The last panel of Figures \ref{fig4} and \ref{fig5} displays the PSF fit and removal of the white 
dwarf at the expected photospheric level; a point-like residual (excess) is seen at the location 
of the white dwarf.  

To assess the potential flux contamination by the galaxy, a radial profile and vector analysis of 
the white dwarf flux was performed.  Both 1D gaussian profile fitting and a raw vector cut (each
of 3\,pixel width) were executed across the white dwarf in the direction towards and away from 
the galaxy, and the fluxes on either side were compared.  From these analyses, it is found that
at $r=2$\,pixels from the white dwarf centroid, the side nearest the galaxy contains at most 15\% 
more flux than the side opposite at 3.6\,$\mu$m, and no more than 10\% at the three longer 
wavelengths.  Assuming a worst case scenario where one-half of the photometric aperture is 
biased in this manner -- certain not to be the case -- then the galaxy contributes a maximum 
of 5\% of the measured flux at all wavelengths except 3.6\,$\mu$m, where it is limited to 7.5\%.
The excess flux (measured$-$photosphere) relative to the measured flux is 27, 32, 39, and 55\%
at 3.6, 4.5, 5.7, and 7.9\,$\mu$m, and hence the potential influence of the galaxy cannot account
this excess.  From this analysis, it appears that HE\,0106$-$3253 has an infrared excess.

{\em HS\,0307$+$0746}.  Figure \ref{fig1} reveals an infrared excess around this star based 
on the plotted 10\,500\,K DA model and the short wavelength photometry.  The $HK$ data 
are anchored to the 2MASS $J$-band measurement of $16.39\pm0.14$\,mag, where the $J-H$ 
and $H-K$ colors were derived spectroscopically by \citet{kil06}.  The only optical, photoelectric 
photometry comes from the Carlsberg Meridian Catalog (CMC) where two measurements are 
given as $r'=16.48\pm0.08$\,mag.  \citet{koe06} give $T_{\rm eff}=10\,200$\,K and $\log\,g=7.9$ 
for the white dwarf, similar to the plotted model that agrees well with the {\em GALEX} far- and 
near-ultraviolet fluxes.  If the model used here is sufficiently accurate, the putative excess at 
HS\,0307$+$0746 appears at the $4-6\sigma$ level at all IRAC wavelengths.

{\em Ton\,345}.  The {\em Spitzer} data for this star is analyzed elsewhere, but the {\em AKARI} 
IRC fluxes reported here are original and appear broadly consistent with the near-infrared and 
{\em Spitzer} IRAC fluxes \citep{mel10,bri10}.

{\em PG\,1225$-$079 and GD\,303}.  The 7.9\,$\mu$m IRAC flux at PG\,1225$-$079 
indicates a $4\sigma$ excess (Figure \ref{fig2}) while all other photometric data are consistent 
with photospheric emission.  If real, this star joins G166-58 as a metal-rich white dwarf with a 
long wavelength IRAC excess but no short wavelength excess; these two objects may form a 
class of object whose orbiting disks have large inner holes.  GD\,303 may have a marginal 
photometric excess at the $2\sigma$ level (Figure \ref{fig2}).

\begin{figure}
\epsscale{1.1}
\plotone{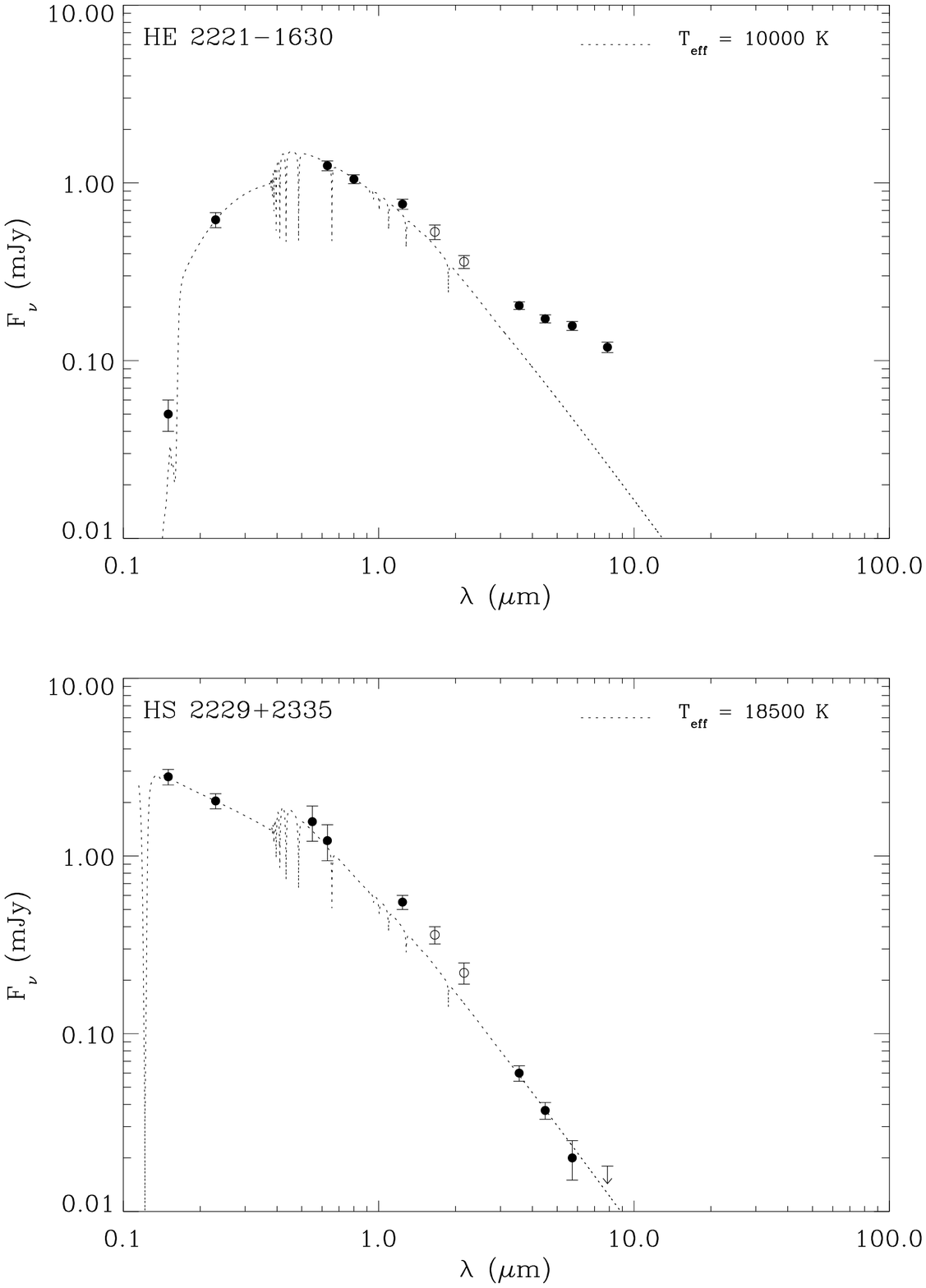}
\caption{SEDs of HE\,2221$-$1630 and HS\,2229$+$2335 fitted with DA white dwarf 
models \citep{koe09b}.  Filled circles represent the photometry \citep{cmc06,skr06,sts06,
den05,zac05,mar05,mon03}, while open circles are fluxes based on spectroscopically 
derived colors \citep{kil06}, and hence not independent measurements.The downward 
arrow is a $3\sigma$ upper limit.\\
\label{fig3}}
\epsscale{1.0}
\end{figure}

{\em HE\,2221$-$1630}.  This metal-polluted white dwarf shows the clearest mid-infrared 
excess (Figure \ref{fig3}) of all the science targets.  \citet{koe06} give $T_{\rm eff}=10\,100$\,K 
for this star.

\subsection{Notes on Stars Without Infrared Excess}

{\em GD\,190}.  The 2MASS $K_s=14.96\pm0.13$\,mag measurement (Figure \ref{fig2}) 
for this helium- and carbon-rich white dwarf appears erroneous at the $2.5\sigma$ level.

{\em GD\,205}.  The fit to the spectral energy distribution (SED) shown in Figure \ref{fig2} 
employs a higher $T_{\rm eff}$ than that found by previous studies; 18\,500\,K \citep{vos07},
19\,700\,K \citep{koe05}.  However, the 20\,500\,K model better matches all the photometric
data, including the {\em GALEX} fluxes.

{\em 2229$+$235}.  While the native pixel scale IRAC images of this star reveal a single 
source (single Airy disk) with an elongated core, the $0\farcs6$\,pixel$^{-1}$ mosaics spatially 
resolve a neighboring source at $2\farcs0$ and position angle $336\arcdeg$, as measured by 
{\sf daophot}.  The 2MASS $J$- and $H$-band images of the white dwarf may show a slight 
elongation in the expected direction, but the S/N in these images is too low to be certain 
(especially at $H$-band).  Photometric deconvolution was performed with {\sf daophot}, 
yielding good results at 3.6 and 4.5 $\mu$m; errors of 6\% and 8\% respectively for the white 
dwarf.  While the neighboring source is likely extragalactic in nature, there is no evidence of 
elongation in the IRAC images, and there are no significant residuals in the PSF subtracted 
images of both sources at either of the shorter wavelengths.

Unfortunately, reliable optical or near-infrared photometry of HS\,2229$+$2335 is lacking; 
the best data are $r'=16.18\pm0.25$\,mag \citep{cmc06}, and $J=16.16\pm0.09$\,mag 
\citep{skr06}.  The flux plotted at $V$-band is the average and standard deviation of five 
photographic plate magnitudes from NOMAD, GSC2.3, and USNO-B1.0 \citep{sts06,zac05,
mon03}.  If interstellar extinction towards the white dwarf is negligible, the {\em GALEX} and 
IRAC data agree with an 18\,500\,K, $\log\,g=8.0$ DA model \citet{koe06}.  Near-infrared 
photometry would better constrain the SED of this white dwarf, but it does not appear to 
have a mid-infrared excess.

\subsection{Disk Models}

\begin{figure}
\epsscale{1.1}
\plotone{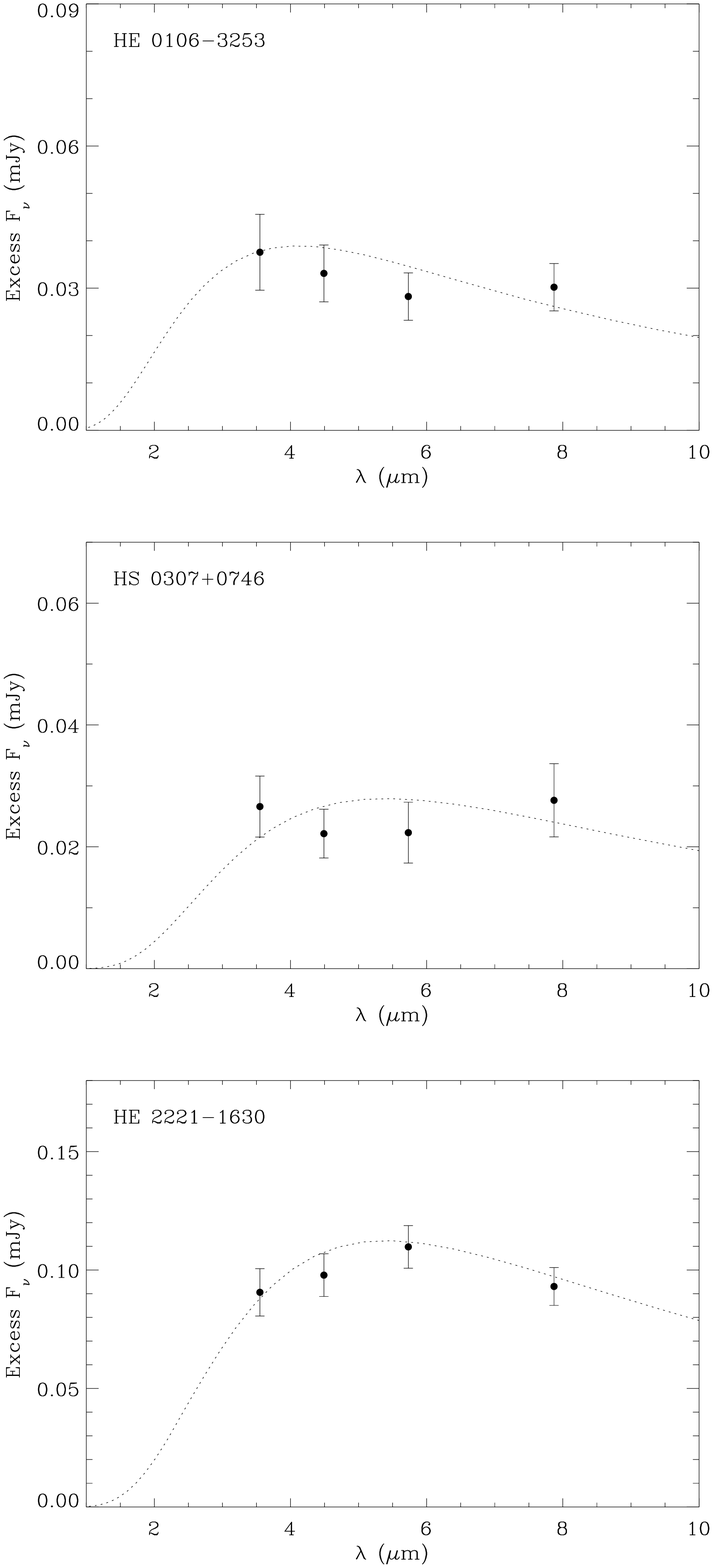}
\caption{The simple flat ring model applied to the excess fluxes at HE\,0106$-$3253, 
HE\,0307$+$0746, and HE\,2221$-$1630.\\
\label{fig7}}
\epsscale{1.0}
\end{figure}

The infrared excesses at HE\,0106$-$3253, HS\,0307$+$0746, and HE\,2221$-$1630 were 
fitted with the circumstellar disk models of \citet{jur03}.  These disks are geometrically thin, flat
rings of dust vertically optically thick at wavelengths up to 20\,$\mu$m.  Figure \ref{fig7} shows 
the model fits, while Table \ref{tbl6} gives the model parameters.  The models do a reasonable 
job of reproducing the disk fluxes.  Both HE\,0106$-$3253 and HS\,0307$+$0746 have a relatively 
high 3.6\,$\mu$m flux relative to the model; this behavior is also seen for the inner disk region at 
PG\,1457$-$086 \citep{far09a}.  It is conceivable that some disk warping in the inner regions is 
responsible for the relatively high observed flux at short infrared wavelengths, similar to that 
seen and modeled for GD\,56 \citep{jur09a,jur07a}.

\section{DISCUSSION}

\subsection{The Existence of Narrow Dust Rings}

\begin{deluxetable*}{cccccccc}
\tabletypesize{\footnotesize}
\tablecaption{White Dwarf Flat Ring Model Parameters \label{tbl6}}
\tablewidth{0pt}
\tablehead{
\colhead{WD}				&
\colhead{$T_{\rm eff}$}		&
\colhead{$R/d$}			&
\colhead{$T_{\rm inner}$}		&
\colhead{$T_{\rm outer}$}		&
\colhead{$r_{\rm inner}$}		&
\colhead{$r_{\rm outer}$}		&
\colhead{$\cos{i}$}			\\			
&(K)
&$(10^{-12})$	
&(K)	
&(K)	
&($R_{\odot}$)	
&($R_{\odot}$)		
&}

\startdata

0106$-$328		&15\,700		&3.4		&1400	&1100	&0.19	&0.27	&0.15\\
0307$+$077		&10\,200		&3.5		&1050	&850	&0.17	&0.22	&0.40\\
2221$-$165		&10\,100		&4.3		&1200	&750	&0.14	&0.27	&0.50

\enddata

\tablecomments{The first column is the input temperature of the central star, followed by the ratio
of the stellar radius to its photometric distance.\\}

\end{deluxetable*}

Two white dwarfs in this study --  HE\,0106$-$3253 and HS\,0307$+$0746 -- appear to have 
narrow ($\Delta r < 0.1\,R_{\odot}$) rings of dust similar to PG\,1457$-$086, when compared 
to the ensemble of white dwarfs with warm circumstellar disks \citep{far09a}.  In fact, their 
disk widths could be even smaller as the models presented here are degenerate between 
disk width and inclination.  For example, if the inclination of the disks at HE\,0106$-$3253 
or PG\,1457$-086$ are actually close to face-on ($i=0\arcdeg$), the radial extent of their 
dust is only $0.01\,R_{\odot}$, roughly 1\,$R_{\oplus}$ and 10 times smaller than the rings
of Saturn, yet still over 10 times the diameter of all Solar System asteroids save Ceres.
This finding has an important implication.

The detection of narrow dust rings suggests the asteroid accretion model may apply 
to additional, and potentially many, stars without a clear infrared excess.  Take the most 
modest excess yet detected, that of PG\,1457$-$086: a face-on disk of width 0.01\,$R_
{\odot}$ and producing a 15\% ($3\sigma$) excess would become difficult to confirm above 
$i=50\arcdeg$, corresponding to an excess of only 10\%.  While such disks may appear slight 
when compared to the majority of infrared excesses discovered so far, they could easily contain 
sufficient mass to fuel the observed photospheric pollutions in DAZ and DBZ stars.  At the high 
densities necessary to prevent collisional dust erosion within the disk \citep{far08}, the thin 
ring above could harbor over 10$^{22}$\,g of matter within a disk height of 10\,m, and readily 
supply heavy elements at 10$^{9}$\,g\,s$^{-1}$ for nearly 10$^6$\,yr.  The discovery of three 
relatively narrow rings indicates there may be more subtle excesses awaiting detection.

Table \ref{tbl7} lists a comparison of the infrared excesses at all known white dwarfs with dust.
To calculate the fractional dust continuum luminosity, $\tau=L_{\rm IR}/L_{\rm WD}$, the excess 
flux in the $2-6\,\mu$m region was fitted by a single temperature blackbody, and its bolometric 
flux divided by the stellar flux (similar to those plotted in the SED figures).  There are currently 
three white dwarfs with $T\sim1000$\,K dust whose infrared excesses are less than $10^{-4}$ 
of the incident starlight: HE\,0106$-$3253, WD\,1041$+$091 (SDSS\,J104341.53$+$085558.2), 
and PG\,1457$-$086.  The stars G166-58 and PG\,1225$-$079 owe their modest excesses to 
a lack of $T>500$\,K dust, hence their low $\tau$ are qualitatively different.  It is noteworthy that 
all stars with $\tau>0.02$ had their disks discovered via ground-based observations.

\begin{deluxetable*}{ccccc}
\tabletypesize{\footnotesize}
\tablecaption{Thermal Continuum Excess at White Dwarfs with Dust \label{tbl7}}
\tablewidth{0pt}
\tablehead{
\colhead{WD}						&
\colhead{Name}					&
\colhead{Stellar Model}				&
\colhead{$2-6\,\mu$m Blackbody$^a$}	&
\colhead{$\tau=L_{\rm IR}/L_{\rm WD}$}\\
&
&$T_{\rm WD}$ (K)	
&$T_{\rm IR}$ (K)	
&}

\startdata

0106$-$328	&HE\,0106$-$3253				&16\,000		&1400		&0.0008\\
0146$+$187	&GD\,16						&11\,500		&1000		&0.0141\\
0300$-$013	&GD\,40						&15\,000		&1200		&0.0033\\
0408$-$041	&GD\,56						&14\,500		&1000		&0.0257\\
0307$+$077	&HS\,0307$+$0746				&10\,500		&1200		&0.0028\\
0842$+$231	&Ton\,345						&18\,500		&1300		&0.0048\\
1015$+$161	&PG							&19\,500		&1200		&0.0017\\
1041$+$091	&SDSS\,104341.53$+$085558.2	&18\,500		&1500		&0.0008\\
1116$+$026	&GD\,133						&12\,000		&1000		&0.0047\\
1150$-$153	&EC\,11507$-$1519				&12\,500		&900		&0.0216\\
1226$+$110	&SDSS\,122859.93$+$104032.9	&22\,000		&1000		&0.0042\\
1225$-$079:	&PG							&10\,500		&300:		&0.0005:\\
1455$+$298	&G166-58						&7500		&500		&0.0015\\
1457$-$086	&PG							&20\,000		&1800:		&0.0007\\
1729$+$371	&GD\,362						&10\,500		&900		&0.0235\\
2115$-$560	&LTT\,8452					&9500		&900		&0.0092\\
2221$-$165	&HE\,2221$-$1630				&10\,100		&1000		&0.0076\\
2326$+$049	&G29-38						&11\,500		&1000		&0.0297

\enddata

\tablecomments{Measured infrared excess from thermal continuum emission between 2 and 
6\,$\mu$m, as most stars lack spectroscopic data on any potential silicate emission in the $8-
12\,\mu$m region.}

\tablenotetext{a}{This single temperature is a zeroth order approximation of the disk SED.\\}

\end{deluxetable*}

A disk of radial width 0.01\,$R_{\odot}$ may result from the tidal disruption of an asteroid 
perturbed into an orbit with periastron near 0.2\,$R_{\odot}$, and dust spreading timescales 
suggest the material need not expand to a much greater radius.  In the context of {\em Spitzer} 
searches for dust at polluted white dwarfs, highly attenuated rings of dust may persist at some 
white dwarfs that lack apparent infrared excess.  Without supplementary ground-based, 
near-infrared photometry, the mild infrared excess at PG\,1457$-$086 would not be obvious 
\citep{far09a,far09b}.  A combination of accurate optical and near-infrared photometry with 
appropriate atmospheric models may reveal excesses previously missed.  Ideally, high 
signal-to-noise mid-infrared spectroscopy spanning the 10\,$\mu$m silicate feature should be 
capable of detecting very modest dust emission, such as that seen at HD\,69830 \citep{lis07}.  
Spectroscopy with the Mid-Infrared Instrument for {\em JWST} would be an ideal choice to 
reveal the most subtle infrared excesses.
	
\subsection{A Class of Disks with Enlarged, Dust-Free Inner Regions?}

\begin{figure}
\epsscale{1.1}
\plotone{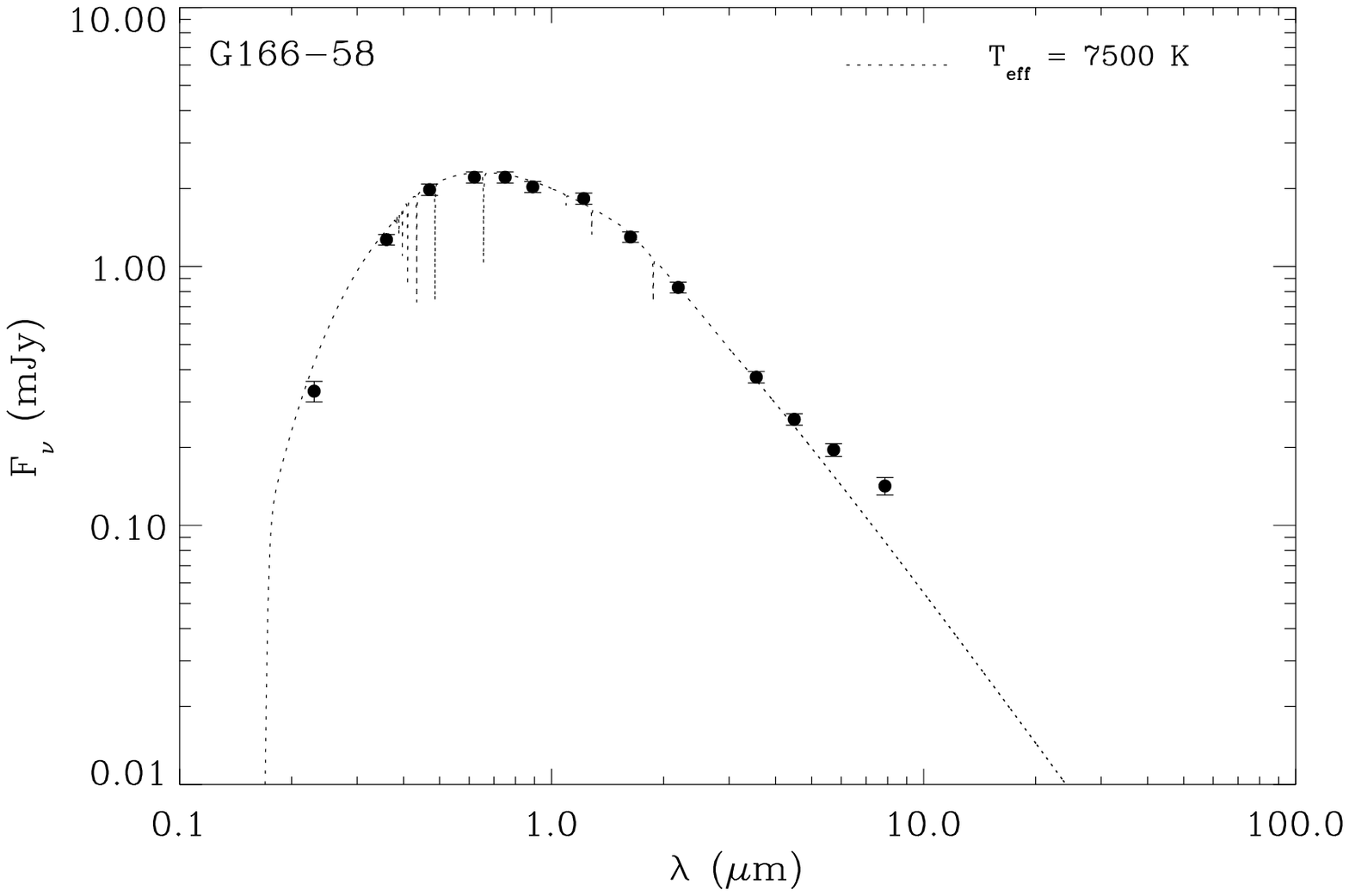}
\caption{SED of G166-58 fitted with a DA white dwarf model 
\citep{koe09b}.  Filled circles represent the photometry \citep{aba09,far09b,mar05}.
\label{fig8}}
\epsscale{1.0}
\end{figure}

Among white dwarfs with circumstellar dust, G166-58 was the first star observed to have 
an beginning long ward of 4\,$\mu$m \citep{far08}.  Figure \ref{fig8} revisits the available 
photometry for this DAZ star and plots the IRAC data, together with IRTF $JHK$, SDSS 
$ugriz$, and {\em GALEX} near-ultraviolet photometry.  These data are fitted with a 7500\,K, 
$\log\,g=8.0$ DA model that reproduces the photometry well to 4.5\,$\mu$m.  The IRAC fluxes 
have changed marginally following analysis of the newly available $0\farcs6$\,pixel$ ^{-1}$ 
mosaics for this star.  The excess can be reproduced by a 500\,K blackbody, implying the 
inner dust disk edge cannot be significantly warmer.  The flat ring model then predicts the 
innermost dust is located at 0.28\,$R_{\odot}$, and far from where silicate grains should 
rapidly sublimate.  Therefore, the dust disk appears to have an enlarged, dust-free inner 
region.

Interestingly, PG\,1225$-$079 appears to have an excess only at 7.9\,$\mu$m and may 
also represent a dust disk with a large inner `hole'.  An excess flux is difficult to confirm 
with the current data alone, yet the IRAC images are of good quality and free of neighboring 
sources.  If the excess is confirmed, the dust-free, inner region at PG\,1225$-$079 should be 
around three times larger than that inferred for G166-58, but still located well within the Roche 
limit for kilometer size or larger solid (rocky or icy) bodies \citep{dav99}.  There are two distinct 
possibilities for such large dust-free zones:  1) the region is dominated by gaseous matter 
emitting inefficiently in the infrared or, 2) the region is depleted relative to the dust inferred 
from the observed infrared excess.

The first possibility can arise if the inner dust disk region is impacted by a moderate 
size asteroid whose solid mass and angular momentum is dissipated via collisions prior to 
vaporizing the entire disk \citep{jur08}.  In Saturn's rings, the observed radial dark lanes or 
spokes are the result of meteorite impacts, whose kinetic energies are sufficient to produce 
fine dust, gas, and (importantly) plasma, which interact dynamically within this environment 
to obscure the rings \citep{goe83,mor83}.  Similar impacts upon dust disks at white dwarfs 
should release kinetic energies 1000 times higher due to the difference in gravitational 
potential; the model radii and sizes of dust rings at white dwarfs are very similar to those 
of Saturn. 

This model is attractive because it can account for disks with enlarged, dust-free inner 
regions or small outer radii (including narrow rings) as well as high accretion rate DAZ 
white dwarfs that lack infrared excess out to 7.9\,$\mu$m yet require a nearby, ongoing 
source of photospheric heavy element pollution.  G166-58 has a relatively long diffusion 
timescale for a DAZ star, around 2000\,yr for the observed calcium \citep{koe09a}.  This 
calculated lifetime does not guarantee accretion is ongoing, but the presence of orbiting 
dust makes it likely.

Perhaps more important, the absence of detectable dust need not imply an absence of
accretion and photospheric pollution.  While the disks at G166-58 and PG\,1225$-$079 may
be partly composed of dust particles which are detected in the infrared, and partly of gaseous 
heavy elements which remain unseen, most metal-contaminated white dwarfs lack infrared
dust signatures altogether \citep{far09a}.  In many of these cases accretion must be ongoing
and can be accounted for by the presence of either tenuous dust rings or completely gaseous 
disks \citep{jur08}; unseen brown dwarf companions have been ruled out \citep{far09a,far08}.

The second possibility for dust disks with large inner holes is a relative dearth of dust (and 
gas) due to prior accretion; i.e. a partly consumed disk.  PG\,1225$-$079 has a heavy element 
diffusion timescale typical of cooler DBZ stars, over 1\,Myr for calcium \citep{koe09a}.  In this 
case the lifetime of metals implies that accretion need not currently proceed at a high rate, 
and the observed metals may be residuals from an older event.  Although the presence of 
an infrared excess, and hence the possibility of orbiting dust, make an ongoing accretion 
scenario likely, it raises an important question about the disk depletion timescale at white 
dwarfs.  A prediction of ceased accretion would be higher relative atmospheric abundances
of lighter elements \citep{koe09a}, and is testable with good abundance data for PG\,1225$-$079.

In the appendix, a simple calculation demonstrates that radiation drag (the Poynting-Robertson 
effect) on particles is insufficient to drive the infall of disk material and account for the distribution 
of inferred heavy element accretion rates at disk-polluted DAZ stars \citep{far09a,jur07b,koe06}.  
At the inner disk edge, viscous spreading due to collisions between differentially rotating ring
annuli and random motions among solids \citep{esp93} must be boosted by the presence of 
gas (via sublimated or collisionally vaporized dust); such enhancement must be significant.  
\citet{far08} have shown that grains within white dwarf disks will collide on a timescale at 
least an order of magnitude shorter than their Poynting-Robertson lifetime.  Therefore, high 
viscosity resulting from dust-gas interactions must be ultimately responsible for the implied 
accretion rates from observed disks.  For a disk that is otherwise optically thick at all relevant 
wavelengths, only a tiny fraction of its total volume -- only the inner edge -- will be illuminated 
by the central star.  There is no compelling reason why the remainder of the disk should not 
evolve as a particulate ring, having a very low dust viscosity \citep{esp93}, and be gradually
whittled away via gas drag at the inner edge.

\subsection{The Emerging Picture}

\begin{figure}
\epsscale{1.1}
\plotone{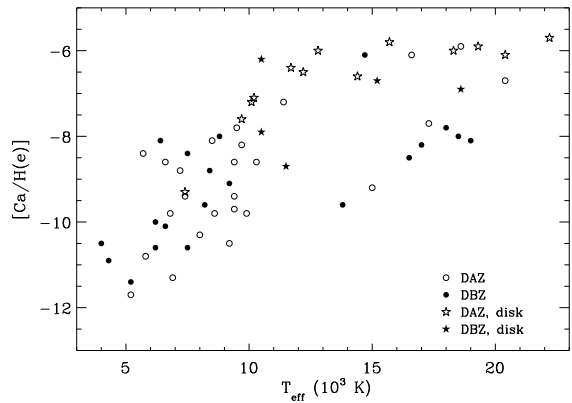}
\caption{Measured calcium abundance versus effective temperature for all 61 metal-rich 
white dwarfs observed with {\em Spitzer} IRAC.  Abundances are expressed as  [X/Y] = 
$\log\,[n({\rm X})/n({\rm Y})]$.
\label{fig9}}
\epsscale{1.0}
\end{figure}

\begin{figure}
\epsscale{1.1}
\plotone{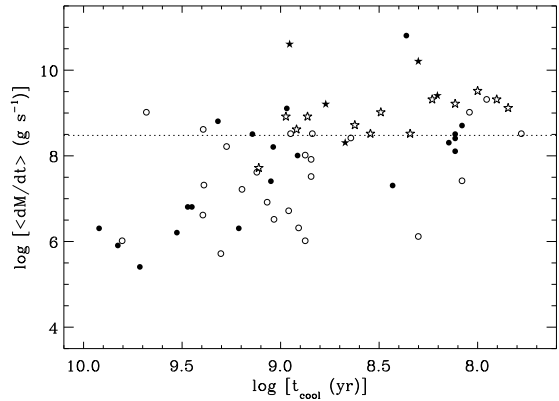}
\caption{Time-averaged dust accretion rates versus cooling age for all 61 metal-rich white 
dwarfs observed with {\em Spitzer} IRAC.  The dotted line represents an accretion rate of
$3\times10^8$\,g\,s$^{-1}$.  Table \ref{tbl8} lists the ($x,y$) coordinates for the new dust 
disk identifications.\\
\label{fig10}}
\epsscale{1.0}
\end{figure}

Figures \ref{fig9} and \ref{fig10} plot the metal abundances versus effective temperature 
and time-averaged metal accretion rate versus cooling age for 60 {\em Spitzer} observed 
metal-rich white dwarfs.  All accretion rates were re-calculated following \citealt{far09a} 
using recent results on metal diffusion timescales and convective envelope masses for both 
DAZ and DBZ white dwarfs, including mixed cases (\citealt{koe09a}; D. Koester 2009, private 
communication); parameters for the newly observed Cycle 5 stars are listed in Table \ref{tbl8}.
Plotted as newly discovered disks are HE\,0106$-$3253, HE\,0307$+$0746, HE\,2221$-$1630, 
and tentatively PG\,1225$-$079.  To these are added three white dwarfs orbited by both dust 
and gaseous metals (\citealt{bri09}; C. Brinkworth 2008, private communication), while 
PG\, 1632$+$177 (observed by \citealt{far08} and listed as DAZ in \citealt{far09a}) has 
been removed from this growing sample, as it is not metal-lined \citep{zuc03}.

The new additions to the plots continue the trend that metal accretion rates $dM/dt \ga 3\times 
10^8$\,g\,s$^{-1}$ are those that are most likely to exhibit infrared excess \citep{far09a}; the sole 
exception is G166-58.  Two of the disks discovered in this work orbit relatively cool white dwarfs
with cooling ages of 0.73 and 0.94\,Gyr.  These are somewhat older than the previous average 
cooling age of white dwarfs with disks as found in \citet{far09a}, and may indicate relative 
longevity for orbiting dust.

Viscous spreading among solids may allow optically thick dust rings at white dwarfs to 
persist for at least 10\,Myr timescales, possibly orders of magnitude longer, analogous to 
planetary rings.  The inner radius of Saturn's rings represents the furthest inner extent of its 
spreading over a period up to a few Gyr \citep{esp08}.  For dust rings at white dwarfs, gas 
drag should decrease the viscous timescale by orders of magnitude, but the gas content in 
most white dwarf disks is presently unconstrained.  At the inner edge of a typical dust disk, 
rapid sublimation of dust will produce significant gas, as may grain-grain collisions.  Such 
viscosity enhancements are necessary to reproduce the dust accretion rates inferred for
stars with short photospheric metal lifetimes.

Dust rings at white dwarfs are unlikely to be as simple as modeled here; calcium triplet 
emission lines at three metal-polluted stars can be approximately reproduced by orbiting 
disks, but there are asymmetries which indicate deviations from the model \citep{gan08}.  
Yet it is noteworthy that the simple model works rather well given only three free parameters,
two of which -- the inner and outer radius -- are physically well-motivated.

Significant orbital energy must be shed to transform an asteroid on a perturbed, eccentric 
orbit into a disk contained entirely within the Roche limit of a white dwarf. The primordial 
energy may dissipate partially via collisions which initially widen the disk (or perhaps not 
in the case of narrow rings), yet an efficient way to discard orbital energy is within a large 
fragment of the tidally disrupted parent body.  Surviving fragments may be responsible for 
warps invoked in some disk models, such as GD\,362 and GD\,56 \citep{jur09a,jur07b}.  
Saturn's ring-moon Daphnis -- 8\,km in diameter and comparable to a small asteroid -- causes 
1\,km vertical disturbances in the A ring; 100 times larger than the height of the unperturbed 
rings \citep{wei09}.  The more substantial Mimas, lying just outside Saturn's main rings (and 
its Roche limit) with a mass nearly identical to the total mass of the rings, is responsible for the 
Cassini Division between the A and B rings.  A Mimas-analog of 10$^{22}$\,g could orbit a 
white dwarf beyond 2.0\,$R_{\odot}$ and not be tidally destroyed.  Such a body with an 
inclined (relative to the disk) or eccentric orbit, could easily excite warping or eccentricity 
within a white dwarf dust ring \citep{bor84}.  Dynamical modeling of tidal disruption events 
may shed light on the frequency, size, and orbital distribution of residual fragments.

\begin{deluxetable*}{ccccccccc}
\tabletypesize{\footnotesize}
\tablecaption{Metal-Polluted White Dwarf Target Parameters \label{tbl8}}
\tablewidth{0pt}
\tablehead{
\colhead{WD}					&
\colhead{SpT}					&
\colhead{$M$}					&
\colhead{$T_{\rm eff}$}			&
\colhead{$t_{\rm cool}$}			&
\colhead{[H/He]}				&
\colhead{log $<dM_{\rm z}/dt>$}	&
\colhead{log ($M_{\rm z}$)}		\\
&
&($M_{\odot}$)	
&(K)				
&(Gyr)				
&					
&(g\,s$^{-1}$)	
&(g)}

\startdata

0047$+$190	&DAZ	&0.51	&16\,600	&0.11	&\nodata				&9.0		&\nodata\\
0106$-$328	&DAZ	&0.62	&15\,700	&0.17	&\nodata				&9.3		&\nodata\\
0307$+$077	&DAZ	&0.66	&10\,200	&0.73	&\nodata				&8.9		&\nodata\\
0842$+$231	&DBZ	&0.71	&18\,600	&0.16	&$-4.5$\tablenotemark{a}	&9.4		&21.7\\	
1011$+$570	&DBZ	&0.60	&18\,000	&0.12	&$-5.5$\tablenotemark{a}	&8.7		&21.2\\
1225$-$079	&DZA	&0.58	&10\,500	&0.59	&$-4.0$				&9.2		&23.1\\
1709$+$230	&DBAZ	&0.65	&18\,500	&0.13	&$-4.4$				&8.4		&20.7\\
2221$-$165	&DAZ	&0.73	&10\,100	&0.94	&\nodata				&8.9		&\nodata\\
2229$+$235	&DAZ	&0.57	&18\,600	&0.09	&\nodata				&9.3		&\nodata

\enddata

\tablecomments{The seventh column lists time-averaged metal infall rates, calculated 
assuming a steady state between accretion and diffusion.  The eighth column shows the 
mass of metals in the convective envelopes of the DBZ stars, based on the observed calcium 
abundances and assuming this represents 1.6\% of the total mass of heavy elements, as in
the bulk Earth \citep{all95}.}

\tablenotetext{a}{Upper limit.\\}

\end{deluxetable*}

\section{CONCLUSIONS}

Three (possibly four) newly identified dust disks orbiting metal-polluted white dwarfs are
discovered by {\em Spitzer} IRAC as mid-infrared excess emission over $3-8\,\mu$m.  The 
fractional infrared (dust continuum) luminosity from HE\,2221$-$1630 is relatively large and 
around 0.008.  In contrast, HE\,0106$-$3253 and HE\,0307$+$0746 with their much smaller 
infrared luminosities are modeled by narrow rings of orbiting dust.  Even narrower disks at 
white dwarfs are possible, and perhaps likely given the necessity for ongoing heavy element 
pollution at the bulk of DAZ stars.  An attenuated, flat disk as narrow as half an earth radius 
can harbor up to $10^{22}$\,g of material and be sufficient to reproduce the inferred 1) metal 
accretion rates of DAZ stars for at least 10$^4$\,yr, and 2) mass of heavy elements in the 
convective envelopes of DBZ stars.  Yet such a tenuous ring of dust would not exhibit an 
infrared excess at the 10\% level for any inclination.  The only evidence of accretion may 
be the photospheric pollution; other signatures such as X-ray emission seen in accreting 
binaries require much higher mass infall rates than inferred for metal-enriched white dwarfs.
Therefore, elemental abundances for polluted stars without an infrared excess are still likely 
to represent the bulk composition of accreted minor planets.

PG\,1225$-$079 is a tantalizing potential disk that emits only at 7.9\,$\mu$m, but which 
may confirm a previously suspected class of dust ring with a large inner hole.  While the 
excess emission from G166-58 is somewhat similar in temperature, a flat ring model 
predicts the inner hole at PG\,1225$-$079 would be significantly larger and may 
represent a disk that is near to being fully accreted; forthcoming data may test this 
scenario.

\acknowledgments

The authors wish to thank the referee F. Mullally for a helpful report which improved the quality of 
the manuscript, and D. Koester for the generous use of his white dwarf model atmospheres.  J. Farihi 
thanks L. Esposito for a brief but illuminating conversation on the dynamical diversity of Saturn's rings 
and moons.  This work is based in part on observations made with the {\em Spitzer Space Telescope}, 
which is operated by the Jet Propulsion Laboratory, California Institute of Technology under a contract 
with NASA, and on observations made with {\em AKARI}, a JAXA project with the participation of ESA.  
J.-E. Lee acknowledges support by the National Research Foundation of Korea grants funded by the 
Korean Government (Nos. 2009-0062865, R01-2007-000-20336-0).  Partial support for this work was 
provided by NASA through an award issued by JPL / Caltech to UCLA, and by the NSF.  

{\em Facility:}	\facility{{\em Spitzer} (IRAC)}; \facility{{\em AKARI} (IRC)}

\appendix

\section{Accretion Rates from Poynting-Robertson Drag on Dust Particles}

\citet{jur07a} demonstrate that dust captured near the Bondi-Hoyle radius (i.e. interstellar 
grains), and accreted under the influence of Poynting-Robertson (PR) drag, cannot account 
for the detected (and upper limit) infrared emission at DAZ white dwarfs.  That is, assuming
metal accretion rates sufficient to reproduce the photospheric metals, the predicted 24\,$\mu
$m emission from grains spiraling inward under radiation drag is typically tens to hundreds 
of times higher than that measured at polluted white dwarfs \citep{far09a,jur07a}.

Here, this argument is extended to consider PR drag on closely orbiting dust rings.  To 
better understand the lifetime of such dust disks in the presence of accretion, consider the 
following simplistic but instructive model.  A flat ring of dust extends outward from 0.2\,$R
_{\odot}$, with thickness $h=10$\,m (typical of Saturn's rings), and consisting of $\rho=2.5$
\,g\,cm$^{-3}$ dust grains.  The ring orbits a compact star with $L=10^{-2}$ $L_{\odot}$, a 
value typical for a 16\,000\,K DA white dwarf \citep{fon01}, whose full starlight penetrates 
the inner disk edge to a depth of one mean free path ($\tau=1$).  The PR lifetime of particles 
at the inner surface is given by \citep{bur79}:

\begin{equation}
t_{\rm pr} = \frac{4 \pi}{3} \frac{c^2 r^2 \rho a}{L Q_{\rm pr}} 
\end{equation}

\medskip\noindent
where $r$ is the distance to the inner disk edge, $a$ is the particle size, and $Q_{\rm pr}$
is the radiative coupling coefficient.  For particles around 1\,$\mu$m or larger, geometrical 
optics is valid and $Q_{\rm pr}=1$.  Applied to a disk of fixed volume and uniform density, the 
shell of mass exposed to the full radiation drag force is

\begin{equation}
M_{\rm pr} = \eta \pi (r'^2 - r^2) h 
\end{equation}

\medskip\noindent
where $\eta$ is the volume mass density of the disk, and $r' - r$ is the depth of the exposed 
shell mass (the mean free path in this case).  For $r' \gg |r'-r|$  One can make the approximation

\begin{equation}
r'^2 - r^2 = (r' - r)(r' + r) \approx 2 l r 
\end{equation}

\medskip\noindent
where $l$ is the mean free path given by

\begin{equation}
l = \frac{1}{\pi a^2 n} 
\end{equation}

\medskip\noindent
and the number density of particles, $n$ is

\begin{equation}
n = \frac{3 \eta}{4 \pi a^3 \rho} 
\end{equation}

\medskip\noindent
Hence the mean free path can be reduced to

\begin{equation}
l=\frac{4 a \rho}{3 \eta} 
\end{equation}

\medskip\noindent
and the exposed shell mass depends only on particle size, and is specifically independent of 
the disk density (or total disk mass in this case)

\begin{equation}
M_{\rm pr} = \frac{8 \pi h r a \rho}{3} 
\end{equation}

\medskip\noindent
Because the PR lifetime of this shell is also proportional to particle size, the mass infall rate at
the inner disk edge is essentially constant for the case of geometric optics

\begin{equation}
\dot M = \frac{M_{\rm pr}}{t_{\rm pr}}=\frac{2 h L Q_{\rm pr}}{c^2 r} 
\end{equation}

\medskip\noindent
and around 6000\,g\,s$^{-1}$ for the parameters given above.  For silicate particles smaller 
than around 1\,$\mu$m, geometric optics is not appropriate; radiative coupling becomes 
more effective until around 0.3\,$\mu$m, then less effective at smaller radii \citep{art88}.  The 
maximum implied mass infall rate due to PR drag calculated in this manner is $1.0\times
10^4$\,g\,s$^{-1}$ and a few to several orders of magnitude shorter than the inferred metal
accretion rates for DAZ white dwarfs.

It should be noted that this model does not apply to disk particles excepting those at the
innermost edge which are exposed to the full starlight.  However, such particles should 
rapidly sublimate at some radius and are then no longer subject to radiation drag, further 
corroborating that PR forces cannot account for the necessary dust/gas infall rates.


\begin{thebibliography}{}

\bibitem[Abazajian et al.(2009)]{aba09} Abazajian,\,K. N., et al. 2009, \apjs, 182, 543

\bibitem[Artymowicz(1988)]{art88} Artymowicz, P. 1988, \apj, 335, L82

\bibitem[All\`egre et al.(1995)]{all95} Allegre, C. J., Poirier, J. P.,Humler, E., \& Hofmann, A.W. 1995, Earth Planetary Sci. Letters, 4, 515
		
\bibitem[Borderies et al.(1984)]{bor84} Borderies, N., Goldreich, P., \& Tremaine, S. 1984, \apj284, 429
		
\bibitem[Brinkworth et al.(2009)]{bri09} Brinkworth, C. S., G\"ansicke, B. T., Marsh, T. R., Hoard, D. W., \& Tappert, C. 2009, \apj, 696, 1402

\bibitem[Brinkworth et al.(2010)]{bri10} Brinkworth, C. S., et al. 2010, in preparation

\bibitem[Burns et al.(1979)]{bur79} Burns, J. A., Lamy, P. L., \& Soter, S. 1979, Icarus, 40, 1
	
\bibitem[Butler et al.(2006)]{but06} Butler, R. P., et al. 2006, 646, 505

\bibitem[Copenhagen University Observatory(2006)]{cmc06} Copenhagen University Observatory 2006, The Carlsberg Meridian Catalog 14 (Strasbourg: CDS)

\bibitem[Davidsson(1999)]{dav99} Davidsson, B. J. R. 1999, Icarus, 142, 525
					
\bibitem[DENIS Consortium(2005)]{den05} DENIS Consortium 2005, The DENIS Database, $3^{\rm rd}$ Release (Strasbourg: CDS)
			
\bibitem[Esposito(1993)]{esp93} Esposito, L. W. 1993, AREPS, 21, 487

\bibitem[Esposito et al.(2008)]{esp08} Esposito, L. W., Meinke, B. K., Colwell, J. E., Nicholson, P. D., \& Hedman, M. M. 2008, Icarus, 194, 278
			
\bibitem[Farihi(2009)]{far09b} Farihi, J. 2009, \mnras, 398, 2091
			
\bibitem[Farihi et al.(2009)]{far09a} Farihi, J., Jura, M., Zuckerman, B. 2009, ApJ, 694, 805
				
\bibitem[Farihi et al.(2008)]{far08} Farihi, J., Zuckerman, B., \& Becklin, E. E. 2008b, \apj, 674, 431

\bibitem[Fazio et al.(2004)]{faz04} Fazio, G. G., et al. 2004, \apjs, 154, 10

\bibitem[Fontaine et al.(2001)]{fon01} Fontaine, G., Brassard, P., \& Bergeron, P. 2001, \pasp, 113, 409 
		
\bibitem[G\"ansicke et al.(2008)]{gan08} G\"ansicke, B. T., Koester, D., Marsh, T. R., Rebassa-Mansergas, A., \& Southworth J. 2008, \mnras, 391, L103
				
\bibitem[G\"ansicke et al.(2006)]{gan06} G\"ansicke, B. T., Marsh, T. R., Southworth, J., \& Rebassa-Mansergas, A. 2006, Science, 314, 1908

\bibitem[Goertz \& Morfill(1983)]{goe83} Goertz, C.\,K., \& Morfill, G. 1983, Icarus, 53, 219

\bibitem[Hagen et al.(1995)]{hag95} Hagen, H. J., Groote, D., Engels, D., \& Reimers, D. 1995, A\&AS, 111, 195
											
\bibitem[Jura(2003)]{jur03} Jura, M. 2003, \apj, 584, L91

\bibitem[Jura(2008)]{jur08} Jura, M. 2008, \aj, 135, 1785

\bibitem[Jura et al.(2007a)]{jur07a} Jura, M., Farihi, J., \& Zuckerman, B. 2007a, \apj, 663, 1285
		
\bibitem[Jura et al.(2009a)]{jur09a} Jura, M., Farihi, J., \& Zuckerman, B. 2009a, AJ, 137, 3191

\bibitem[Jura et al.(2009b)]{jur09b} Jura, M., Muno, M., Farihi, J., \& Zuckerman, B. 2009b, ApJ, 699 1473

\bibitem[Jura et al.(2007b)]{jur07b} Jura, M., Farihi, J., Zuckerman, B., \& Becklin, E. E. 2007b, \aj, 133, 1927

\bibitem[Kilic et al.(2008)]{kil08} Kilic, M., Farihi, J., Nitta, A., \& Leggett, S.\,K. 2008, \aj, 136, 111
				
\bibitem[Kilic et al.(2006)]{kil06} Kilic, M., von Hippel, T., Leggett, S.\,K., \& Winget, D. E. 2006, \apj, 646, 474 

\bibitem[Klein et al.(2010)]{kle10} Klein, B., Jura, M., Koester, D., Zuckerman, B., \& Melis C. 2010, \apj, in press (arXiv:0912.1422)

\bibitem[Koester(2009a)]{koe09a} Koester, D. 2009a, \aap, 498, 517

\bibitem[Koester(2009b)]{koe09b} Koester, D. 2009b, to appear in Memorie della Societ\`a Astronomica Italiana, based on lectures given at the School of  Astrophysics ``F. Lucchin'', Tarquinia, June 2008 (arXiv:0812.0482)	
		
\bibitem[Koester \& Wilken(2006)]{koe06} Koester, D., \& Wilken, D. 2006, \aap, 453, 1051
	
\bibitem[Koester et al.(2005)]{koe05} Koester, D., Rollenhagen,\,K., Napiwotzki, R., Voss, B., Christlieb, N., Homeier, D., \& Reimers, D. 2005a, \aap, 432, 1025 

\bibitem[Landolt \& Uomoto(2007)]{lan07} Landolt, A. U., \& Uomoto, A.\,K. 2007, \aj, 133, 768
		
\bibitem[Lisse et al.(2007)]{lis07} Lisse, C. M., Beichman, C. A., Bryden, G., \& Wyatt, M. C. 2007, \apj, 658, 584
		
\bibitem[Martin et al.(2005)]{mar05} Martin, D. C., et al. 2005, \apj, 619, L1
					
\bibitem[McCook \& Sion(2008)]{mcc08} McCook, G. P., \& Sion, E. M. 2008, Catalog of Spectroscopically Identified White Dwarfs (Strasbourg: CDS)

\bibitem[McCook \& Sion(1999)]{mcc99} McCook, G. P., \& Sion, E. M. 1999, \apjs, 121, 1

\bibitem[Melis et al.(2010)]{mel10} Melis, C., Jura, M., Albert, L., Klein, B., \& Zuckerman, B. 2010, \apj, submitted

\bibitem[Monet et al.(2003)]{mon03} Monet, D., et al. 2003, \aj, 125, 984

\bibitem[Morfill \& Goertz(1983)]{mor83} Morfill, G. E., \& Goertz, C.\,K. 1983, Icarus, 55, 111

\bibitem[Murakami et al.(2007)]{mur07} Murakami, H., et al. 2007, \pasj, 59, 369

\bibitem[Onaka et al.(2007)]{ona07} Onaka, T. et al. 2007, \pasj, 59, 401
 	
\bibitem[Petitclerc et al.(2005)]{pet05} Petitclerc, N., Wesemael, F., Kruk, J. W., Chayer, P., Billeres, M. 2005, \apj, 624, 31
		
\bibitem[Skrutskie et al.(2006)]{skr06} Skrutskie, M. F., et al. 2006, \aj, 131, 1163

\bibitem[Space Telescope Science Institute(2006)]{sts06} Space Telescope Science Institute 2006, The Guide Star Catalog Version 2.3, (Baltimore: STScI)

\bibitem[Spitzer Science Center(2006)]{ssc06} Spitzer Science Center. 2006, IRAC Data Handbook Version 3.0 (Pasadena: SSC)
				
\bibitem[Voss et al.(2007)]{vos07} Voss, B., Koester, D., Napiwotzki, R., Christlieb, N., \& Reimers, D. 2007, \aap, 470, 1079

\bibitem[Werner et al.(2004)]{wer04} Werner, M. W., et al. 2004, \apjs, 154, 1

\bibitem[Weiss et al.(2009)]{wei09} Weiss, J. W., Porco, C. C., \& Tiscareno, M. S. 2009, \aj, 138, 272
	
\bibitem[Wisotzki et al.(1996)]{wis96} Wisotzki, L., Koehler, T., Groote, D., \& Reimers, D. 1996, A\&AS, 115, 227

\bibitem[Wolff et al.(2002)]{wol02} Wolff, B., Koester, D., \& Liebert, J. 2002, \aap, 385, 995

\bibitem[Zacharias et al.(2005)]{zac05} Zacharias, N., Monet, D. G., Levine, S. E., Urban, S. E., Gaume, R., Wycoff, G. L. 2005, The NOMAD Catalog (Strasbourg: CDS)
	
\bibitem[Zuckerman et al.(2007)]{zuc07} Zuckerman, B., Koester, D., Melis, C., Hansen, B. M. S., \& Jura, M. 2007, \apj, 671, 872 
						
\bibitem[Zuckerman et al.(2003)]{zuc03} Zuckerman, B., Koester, D., Reid, I. N., \& H\"unsch, M. 2003, \apj, 596, 477
	
\end{thebibliography}
\end{document}